\documentclass[aps,pra,english,showpacs,twocolumn]{revtex4}
\usepackage[english]{babel}
\usepackage[ansinew]{inputenc}
\usepackage{amsfonts}
\usepackage{latexsym}
\usepackage{amsmath}
\usepackage{graphicx}

\begin{document}

\title{Molecule Formation in Optical Lattice Wells by Resonantly Modulated Magnetic Fields}

\author{Jesper Fevre Bertelsen}
\author{Klaus M\o lmer}
\affiliation{Danish National Research Foundation Center for Quantum Optics\\Department of
Physics and Astronomy, University of Aarhus\\DK 8000 Aarhus C, Denmark}

\date{\today}

\begin{abstract}
We present a theoretical model for formation of molecules in an optical lattice well where
a resonant coupling of atomic and molecular states is provided by small oscillations of a
magnetic field in the vicinity of a Feshbach resonance. As opposed to an adiabatic sweep over the full resonance, this provides a coherent coupling with a frequency that can be tuned to meet resonance conditions in the system. The effective Rabi frequencies for this coupling are calculated and simulations show perfect Rabi oscillations. Robust production of molecules with an adiabatic sweep of the modulation frequency is demonstrated. For very large oscillation amplitudes, the Rabi
oscillations are distorted but still effective and fast association is possible.
\end{abstract}

\maketitle

Creation of molecules from ultracold atoms has been successfully achieved by making
an adiabatic sweep of the magnetic field across a Feshbach resonance
\cite{Hodby,Herbig,Xu,Jin,Jochim,RempeMol}. Feshbach resonances can be used to create molecules,
because of the transformation of a scattering state into a new bound state, when the scattering length changes sign. This occurs when the magnetic field varies near a Feshbach resonance \cite{Moerdijk}:
\begin{align}\label{aB}
a_\textrm{sc}(B)=a_\textrm{bg}\left(1-\frac{\Delta}{B-B_0}\right).
\end{align}
Here $a_\textrm{bg}$ is the background value of the scattering length, and $B_0$ and
$\Delta$ are the location and width of the resonance. In a recent publication \cite{Wieman}
it was demonstrated that near Feshbach resonances, molecules can be created by harmonic
modulation of the magnetic field:
\begin{align}\label{field_osc}
B(t)=B'+b\sin(\omega_Bt),
\end{align}
where $\omega_B$ is resonant with the energy difference between
the atomic and the molecular state at the magnetic field $B'$. It is the purpose of this article to analyze this process for two atoms in an optical lattice well, which is a very clean system for which many-body effects can be neglected. One argument in favour of producing the molecules in the lattice resonantly rather than by sweeping across the Feshbach resonance is that it enables the preparation of well-controlled coherent superposition states of atomic and  molecular components.

Ultracold molecules in optical lattices have been created by photoassociation \cite{BlochMol,Ryu}, and proposals for creating heteronuclear molecules have also been presented \cite{DipSup}. In \cite{Ryu}, Rabi oscillations have been seen with this method. In \cite{Esslinger1D,Pitaevskii} the production of molecules via Feshbach resonances in a 1D confining lattice potential has been studied and very recently the first successful creation of molecules in separate wells of a 3-dimensional optical lattice by sweeping across a Feshbach resonance has been reported \cite{Grimm}. As it turns out these molecules can be very long-lived (up to 700 ms) since they do not suffer from decay due to inelastic collisions.
\begin{figure}[!htb]
\begin{center}
\includegraphics[width=0.80\columnwidth]{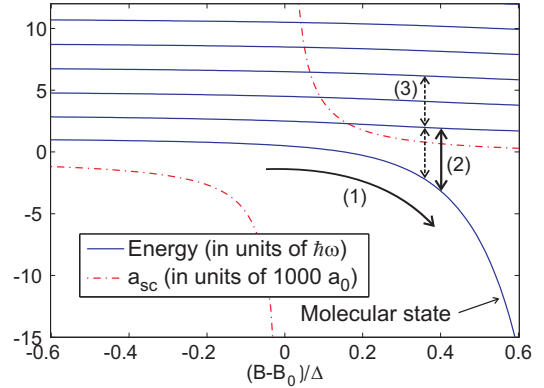}
\caption{\small The \emph{s}-wave spectrum of the potential (\ref{RegDelHam}) around the
$B_0=155\textrm{ G}$ Feshbach resonance in $^{85}$Rb in a harmonic oscillator trap with
$\omega=2\pi\cdot$ 30 kHz. The variation of the scattering length $a_\textrm{sc}$ as a function of
B is shown as a dot-dashed line. Note that the bound (molecular) state with negative energy
occurs in the $a>0$ (repulsive effective interaction) regime. The typical adiabatic
production of molecules by a magnetic field scan across the resonance is indicated by the
arrow (1). Arrow (2) illustrates the production of molecules by resonant association as
analyzed in the present article, while the dotted arrows (3) show an accidental three-level
resonance (discussed in the text).\label{EB}}
\end{center}
\end{figure}

In \cite{Wieman} a commonly used resonance in $^{85}$Rb is applied. It has the parameters \cite{Claussen} $B_0$=155.041 G, $\Delta=$10.71 G, and $a_\textrm{bg}=-443\ a_0$, where $a_0$ is the Bohr
radius. This resonance has previously been used to produce molecules \cite{Hodby} and to
tune the scattering length to make stable atomic BECs of $^{85}$Rb atoms \cite{Cornish}.

We now look at a simple two-atom model for molecule formation in a three-dimensional optical lattice which is sufficiently deep that the atoms do not tunnel
between lattice sites. The system may either be in the Mott insulating state \cite{Mott},
or we may have a mean field state with number fluctuations in the site occupancy, in which
case the collisional interactions apply to the two-atom component of the state on each site
\cite{EsslingerKM}. The assumption is that we have two identical atoms in a lattice well
and neglect couplings to atoms in other wells. The lattice potential is $V(x,y,z)=V_0\sum_{i=x,y,z}\sin^2(kx_i),\qquad k=\frac{2\pi}{\lambda}$, where $\lambda$ is
the wavelength of the standing wave laser field of the lattice, and we make the
harmonic and isotropic approximation in each well:
$V(x,y,z)\simeq\frac{1}{2}m\omega^2\sum_{i=x,y,z}x_i^2$. The potential for two atoms in
this harmonic potential decouples into parts describing the center-of-mass and the relative motion. The
interaction between atoms is modelled by the Huang s-wave pseudopotential \cite{Huang}
and the total potential for the relative motion becomes
\begin{align}\label{RegDelHam}
V(\mathbf r)=\frac{1}{2}(m/2)\omega^2
r^2+\frac{4\pi\hbar^2a_\textrm{sc}(B)}{m}\delta^{(3)}(\mathbf r)\frac{\partial}{\partial r}r
\end{align}
This is an adequate approximation to the more comprehensive analysis \cite{Bolda} which takes into account the spatial confinement and the energy dependence of the scattering length. We note that a more precise treatment amounts in principle to the application of the exact energy eigenvalues and matrix elements instead of the analytical expressions offered by the pseudo-potential in (\ref{RegDelHam}). Such an analysis will lead to the same physical behaviour of the system. For our examples with Rb atoms and typical optical lattice parameters, the characteristic values of the scattering length is about one order of magnitude smaller than the harmonic oscillator length, and we expect that the analytical model provides a reasonable approximation to the exact dynamics \cite{Tiesinga}.

We adopt harmonic oscillator units $\sqrt{\hbar/(m/2)\omega}$ for length and
$\hbar\omega$ for energy, and we express the energy of the s-wave states by a generalized
harmonic oscillator quantum number: $E_i(B) = 2\nu_i(B)+3/2$. Following
\cite{Busch,Greene}, $\nu_i(B)$ are obtained as the solutions of the equation
\begin{align}\label{Eequation}
\frac{2\Gamma(-\nu_i)}{\Gamma(-\nu_i-1/2)}=\frac{1}{a_\textrm{sc}(B)}
\end{align}
where $\Gamma$ is the gamma function.
The energy spectrum is plotted in Fig. \ref{EB} for the $^{85}$Rb resonance.

Assuming a time dependent magnetic field, the eigenstates depend on $a_\textrm{sc}$ and thus on $t$. We
shall label the states $|\psi_n(t)>$, $n=0,1,2,...$ with increasing energy. They will at
every instant form a complete set. Expanding the wave function $|\psi(t)>$ on these states
we obtain the following equations for the expansion coefficients $c_m$:
\begin{align}\label{diffeq}
i\hbar\frac{d c_m(t)}{dt}
=E_m(t)c_m(t)-i\hbar\sum_{n=0}^{\infty}\left.\frac{\partial O_{mn}(t,t')}{\partial
t'}\right|_{t'=t}c_n(t)
\end{align}
where $O_{mn}(t,t')\equiv<\psi_m(t)|\psi_n(t')>$. Introducing the $\nu$-values $\nu_m$ and
$\nu_n$ corresponding to $\psi_m(t)$ and $\psi_n(t')$ respectively and by following a trick
in \cite{Greene}, or by explicitly evaluating the integral of the wave functions
\cite{Integral_table}, we obtain
\begin{align}\nonumber\label{matrixel}
O_{mn} & (t,t')=\frac{f(\nu_m)f(\nu_n)}{\nu_n-\nu_m}
\\&\nonumber
\cdot\frac{\Gamma(-\nu_m)\Gamma(-\nu_n-\frac{1}{2})-\Gamma(-\nu_n)\Gamma(-\nu_m-\frac{1}{2})}{\sqrt{\Gamma(-\nu_m-\frac{1}{2})\Gamma(-\nu_m)}\cdot\sqrt{\Gamma(-\nu_n-\frac{1}{2})\Gamma(-\nu_n)}}
\\&=
\frac{f(\nu_m)f(\nu_n)}{\nu_n-\nu_m}\left(\sqrt\frac{a_\textrm{sc}(t')}{a_\textrm{sc}(t)}-\sqrt\frac{a_\textrm{sc}(t)}{a_\textrm{sc}(t')}\right)
\end{align}
where
\begin{align}
f(\nu)=\frac{1}{\sqrt{\Psi(-\nu)-\Psi(-\nu-\frac{1}{2})}}
\end{align}
and $\Psi(\nu)\equiv\frac{d}{d\nu}\ln(\Gamma(\nu))$ is the digamma function.

Inserting now the oscillating magnetic field (\ref{field_osc}) in (\ref{diffeq}) we get the
following coupling between the states:
\begin{align}\label{coupling}
i\hbar\frac{d c_m(t)}{d t}=E_m(t)c_m(t)+i\hbar\cos(\omega_B
t)\sum_{n=0}^{\infty}\Omega_{mn}(t)c_n(t)
\end{align}
where the Rabi frequencies $\Omega_{mn}$ are given by
\begin{align}\label{Rabi-formula}
\Omega_{mn}&=\frac{-(b/\Delta)\omega_B\Delta^2}{(B-B_0)(B-B_0-\Delta)}f(\nu_n)^2
\left.\frac{\partial O_{\nu_m,\nu_n'}}{\partial\nu_n'}\right|_{\nu_n'=\nu_n}
\end{align}

\begin{figure}[!htb]
\begin{center}
\includegraphics[width=0.80\columnwidth]{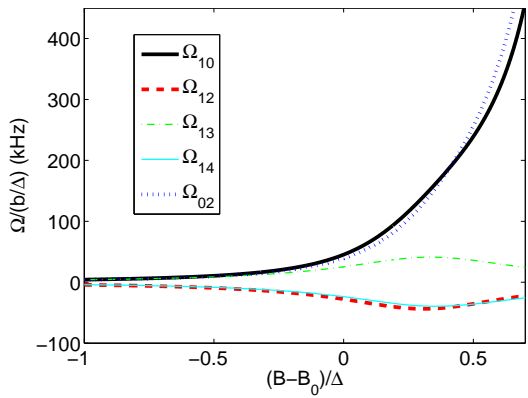}
\caption{\small Amplitude normalized Rabi frequencies on resonance for $^{85}$Rb. The parameters are as
in Fig. \ref{EB}. The states are labelled $0,1,2,\ldots$ with increasing energy. E.g.,
$\Omega_{10}$ is the coupling between the two lowest states. } \label{freq85}
\end{center}
\end{figure}

The Rabi frequencies are proportional to $b/\Delta$, and the rest of the expression is
plotted in Fig. \ref{freq85}. We remark that $\Omega_{nn}=0$. Equivalent to the radiative
coupling of atomic energy levels, if $\hbar\omega_B$ matches the energy difference between
two specific levels $m$ and $n$, Rabi oscillations between these two levels with a constant frequency $\Omega_{mn}$ are expected  as long as $b\ll\Delta$. Notice that due to the factor $\omega_B$ in Eq. \ref{Rabi-formula}, there is a much stronger resonant coupling to the lowest (molecular) state than among the atomic states on the molecular side of the resonance.

Using Eq. (\ref{diffeq}) we have calculated the dynamics of the resonant association
of two $^{85}$Rb atoms in an $\omega=2\pi\cdot 30$ kHz lattice well. Referring to the above
resonance we use $B'=B_0+0.4\Delta=159.33$ G, and $b=0.01$ and we start out in the lowest atomic state (on
the 2$^{nd}$ lowest curve in Fig. \ref{EB}).  The expected resonant frequency is found from
(\ref{Eequation}): $\omega_B^\textrm{res}=(E_1(B')-E_0(B'))/\hbar=5.079\ \omega=2\pi\cdot
152.4$ kHz.
\begin{figure}[!htb]
\begin{center}
\includegraphics[width=0.80\columnwidth]{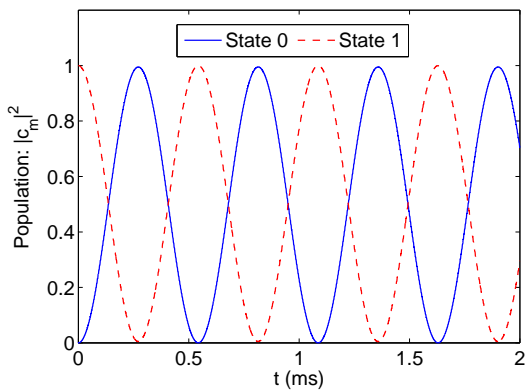}
\caption{\small Dynamics of the resonant process at $B'=B_0+0.4\Delta$. The field is
modulated at an angular frequency $\omega_B=2\pi\cdot 152.4\textrm{ kHz}$, with the
amplitude $b=0.01\Delta$. The other parameters are as in Fig. \ref{EB}.} \label{Rabi30}
\end{center}
\end{figure}
Although the calculation is done with 10 basis states the dynamics in Fig. \ref{Rabi30}. shows full
two-level Rabi-oscillations as one would expect. A fit of the molecular state population to
a $\sin^2(\Omega/2\cdot t)$ function shows that the Rabi frequency is $\Omega=2\pi\cdot
1.84\textrm{ kHz}$ which exactly matches the value of the Rabi frequency displayed in Fig.
\ref{freq85}. Matching the oscillation time such that $\Omega t=\pi$ one
can transform the atomic pair into a molecule with unit probability, and using the interaction time as a control parameter, a wide range of well-controlled superposition states of the molecular and atomic component can also be prepared.

At certain magnetic fields three levels are coupled in a ladder configuration, because
$\hbar\omega_B^\textrm{res}$ matches not only the energy distance to the molecular state,
but also the distance to a higher lying atomic state (see Fig. \ref{EB}). This leads to a more complicated
three-level dynamics, but since the coupling to the higher lying atomic state is much
smaller than the coupling to the molecular state (see Fig. \ref{freq85}), the higher lying
atomic level becomes only sparsely populated. This is fortunate from the point of view of
molecular formation. The lowest B-field on the molecular side of the resonance where such a three-level
resonance occurs is $B'=B_0+0.3582\Delta=158.88$ G shown with the dashed arrows in Fig. \ref{EB} ($\omega_B^\textrm{res}=2\pi\cdot 125.1$ kHz). Simulation of the dynamics in this case
shows oscillations with a peak population of 6 \% in the excited atomic state. All
remaining population is distributed between the molecular and the lowest atomic state. At
the next three-level resonance at $B'=B_0+0.4385\Delta=159.74\textrm{ G}$ the peak
population of the higher lying atomic state becomes only 4 \%.

\begin{figure}[!htb]
\begin{center}
\includegraphics[width=0.8\columnwidth]{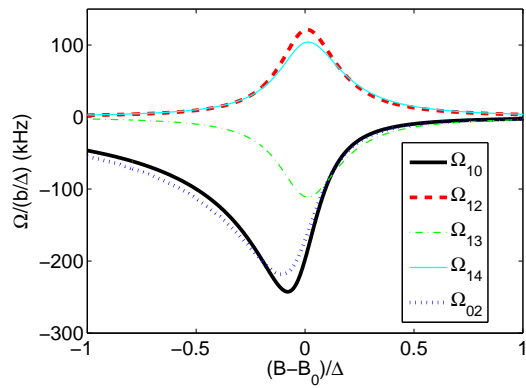}
\caption{\small Amplitude normalized Rabi frequencies in $^{87}$Rb (cf. Fig. \ref{freq85}).
Here the positive scattering length (molecular) side  is to the left of the resonance,
because a$_\textrm{bg}=100.5\ a_0 > 0$. We assume the same trapping frequency
$\omega=2\pi\cdot$ 30 kHz as above.} \label{freq87}
\end{center}
\end{figure}

The scheme should work just as well for Feshbach resonances in other species like $^{87}$Rb although the Feshbach resonances are more narrow and situated at higher magnetic fields. According to \cite{RempeScatt} the broadest resonance in
$^{87}$Rb occurs between atoms in the F=1 m$_\textrm{F}$=1 hyperfine state and has the following parameters: $B_0=$1007.60(3) G, $\Delta=$0.20(3) G, $a_\textrm{bg}=100.5 a_0$. This resonance has been used to produce molecules in an optical dipole trap \cite{RempeMol} and it was also used in the recent optical lattice experiment \cite{Grimm}. Some Rabi frequencies for this system are shown in Fig. \ref{freq87}. The couplings are of the same order of magnitude, but the plot is qualitatively different from the plot in Fig. \ref{freq85}. The difference can be partly attributed to the different behaviour of the scattering length as $B$ approaches $B_0-\Delta$ (positive scattering length side in $^{87}$Rb) compared to when $B$ approaches $B_0+\Delta$ (positive scattering length side in $^{85}$Rb) - see (\ref{aB}). In the latter case the scattering length approaches 0 and the resonant frequency $\omega_B^\textrm{res}$ entering (\ref{Rabi-formula}) diverges for coupling to the molecular state, because the binding energy of the molecular state in the pseudopotential model varies as
\begin{align}
E_0=\frac{-\hbar^2}{2(m/2)a_\textrm{sc}^2}=\frac{-\hbar^2}{ma_\textrm{bg}^2(1-\Delta/(B-B_0))^2}
\end{align}

\begin{figure}[!htb]
\begin{center}
\includegraphics[width=0.8\columnwidth]{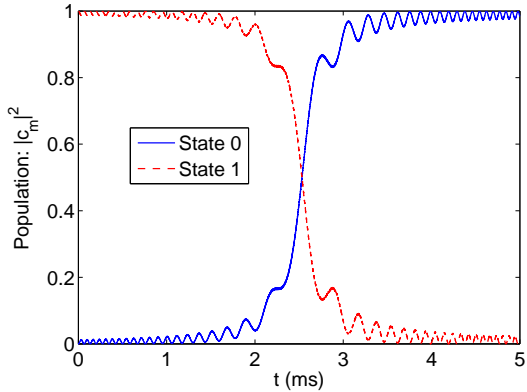}
\caption{\small 5 ms linear frequency sweep from 10\% below to 10\% above the resonance
frequency $\omega_B^\textrm{res}=2\pi\cdot 152.4\textrm{ kHz}$ at an applied field
$B'=B_0+0.4\Delta$. The modulation has the constant amplitude: $b=0.01\Delta$.}
\label{linfreqsweep}
\end{center}
\end{figure}

Application of a $\pi$-pulse in the above scheme relies on the resonance and Rabi
frequencies being exactly known and identical for all lattice sites. A more robust scheme
is to sweep the frequency slowly across the resonance frequency. Using a constant amplitude
linear frequency sweep from 10\% below
 to 10\% above $\omega_B^\textrm{res}$, we get the dynamics shown in Fig.
\ref{linfreqsweep}. After just a few Rabi oscillation periods the molecular state
population settles near 100 \%. In addition to correcting for inhomogeneity effects, a
frequency sweep will also introduce some robustness against technical noise in the magnetic
field value, as long as this noise is present only on longer timescales than that given by
the sweeping time. This is especially interesting regarding applications to more
narrow Feshbach resonances like in $^{87}$Rb.

\begin{figure}[!htb]
\begin{center}
\includegraphics[width=0.90\columnwidth]{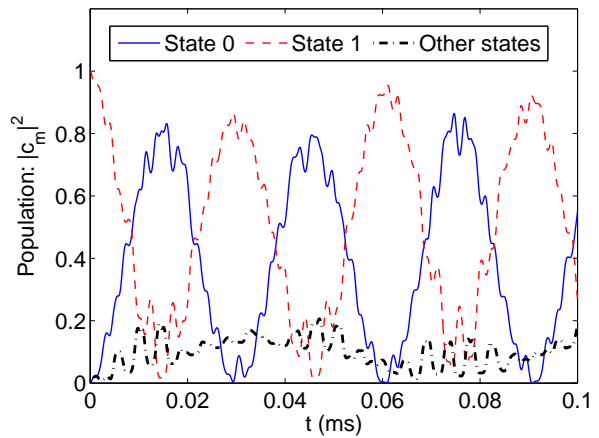}
\caption{\small Dynamics of the process with a very high oscillation amplitude:
$b=0.2\Delta$. $B'=B_0+0.4\Delta$. The resonance frequency has shifted 45 \% compared to
the value in Fig. \ref{Rabi30} so $\omega_B=1.45\cdot 2\pi\cdot 152.4\textrm{ kHz}$. The frequency of the population oscillation is 33 kHz whereas the Rabi frequency predicted from (\ref{Rabi-formula}) is 37 kHz. There are 20 basis states included in the calculation.}
\label{highampl}
\end{center}
\end{figure}

One might also consider the possibility of using larger oscillation amplitudes. This makes the association quicker, can be used to keep the magnitude of the harmonic modulation well above the technical noise level and gives also some robustness against a noisy value of $B'$. Our simulations show that if the coupling strength is not much smaller than $\Delta$, the resonance frequency is
significantly shifted and the oscillations are distorted, but they show a surprisingly high
molecular peak population. For instance with $B'=B_0+0.4\Delta$ and $b=0.2\Delta$ we must
use a value of $\omega_B$ which is 45 \% greater than that in Fig. \ref{Rabi30} to get
maximum population oscillations, and we then get a molecular peak population of around 80 \% (Fig. \ref{highampl}). We attribute the frequency shift to the strongly increasing slope of the molecular level curve (Fig. \ref{EB}) as the B-field is modulated to larger values, farther away from the resonance.

To summarize, we have analyzed the formation of molecules from atoms in an optical lattice
by a resonant modulation of a B-field which causes Rabi-oscillations between the atomic and
the molecular state in the vicinity of a Feshbach resonance. For identical atoms, the
problem separates in a center-of-mass and a relative motion part, and the relevant s-wave
scattering part was solved exactly in terms of known analytical solutions of the
time-independent Schr\"odinger equation. Complete conversion to  molecules was seen both
with a $\pi$-pulse of the interaction and with a frequency swept
interaction. In the case of different atoms, relevant for the formation of
heteronuclear molecules, we expect the coupling between center-of-mass and relative motion
to cause highly non-trivial effects.

In the experiment \cite{Wieman} the large number of $^{85}$Rb atoms is not pairwise confined in a lattice potential, and rather than perfect Rabi oscillations between atoms and molecules, the experiments show small rapid oscillations, dying out towards a steady state with about 30 \% of the atoms converted into molecules. Many-body effects, including collisions in the gas, probably cause the discrepancy with the simple two-atom model. The simple picture remains, however, useful for the identification of the resonant process in the system, and we imagine that the swept modulation frequency, proposed in this article, may lead to enhanced molecular fractions in the many-body system.

Since efficient conversion into molecules inside an optical lattice was recently demonstrated by an adiabatic sweep of the B-field across the full $^{87}$Rb Feshbach resonance \cite{Grimm}, let us conclude by pointing out some relevant features of the modulated field approach. Our approach is equivalent to driven two-level transitions in laser and NMR spectroscopy, and by varying the phase, amplitude and duration of the applied B-field modulation, we have more degrees of freedom to control the transition process. For a number of quantum optics studies, e.g., for quantum non-demolition photon counting and non-classical state preparation and for quantum state tomography, the coherent Rabi oscillation phenomenon is extremely useful, and analogous proposals for trapped atoms \cite{KMprl90} can be implemented with the modulated B-fields. In connection with applications to quantum information, unitary operations are requested that can be applied to unknown initial quantum states, and for this purpose single pulses, or possibly composite pulse sequences with built-in robustness against frequency and amplitude errors, are more versatile than the adiabatic passage across the full resonance.

\acknowledgments{The Authors would like to thank Michael Budde, Nicolai Nygaard, and Uffe Vestergaard Poulsen for comments on the manuscript.}

\end{document}